\documentclass[conference]{IEEEtran}
\IEEEoverridecommandlockouts
\usepackage{cite}
\usepackage{amsmath,amssymb,amsfonts}
\usepackage{algorithmic}
\usepackage{graphicx}
\usepackage{textcomp}
\usepackage{xcolor}
\usepackage{listings}
\usepackage{hyperref}
\usepackage[ruled,vlined,linesnumbered]{algorithm2e}

\usepackage{subfigure}
\usepackage{subcaption}
\usepackage{combelow}

\def\BibTeX{{\rm B\kern-.05em{\sc i\kern-.025em b}\kern-.08em
    T\kern-.1667em\lower.7ex\hbox{E}\kern-.125emX}}
\begin{document}

\title{\textsc{StopHC}: A Harmful Content Detection and Mitigation Architecture for Social Media Platforms}

\author{
    \IEEEauthorblockN{
        Ciprian-Octavian Truic\u{a}$^{1,*}$\thanks{$^*$ These authors contributed equally to this work.},
        Ana-Teodora Constantinescu$^{1,*}$
        and
        Elena-Simona Apostol$^{1,2,*}$   
    }
    \IEEEauthorblockA{
        $^1$ \textit{National University of Science and Technology Politehnica Bucharest, 313 Independen\cb{t}ei, 060042, Bucharest, Romania}\\
        $^2$ \textit{Academy of Romanian Scientists, 3 Ilfov, Bucharest, Romania} \\
        \texttt{
        ciprian.truica@upb.ro,
        aconstantinescu1704@stud.acs.upb.ro,
        elena.apostol@upb.ro
        }
    }
}

\maketitle

\begin{abstract}
The mental health of social media users has started more and more to be put at risk by harmful, hateful, and offensive content.
In this paper, we propose \textsc{StopHC}, a harmful content detection and mitigation architecture for social media platforms. 
Our aim with \textsc{StopHC} is to create more secure online environments.
Our solution contains two modules, one that employs deep neural network architecture for harmful content detection, and one that uses a network immunization algorithm to block toxic nodes and stop the spread of harmful content.
The efficacy of our solution is demonstrated by experiments conducted on two real-world datasets. 
\end{abstract}


\section{Introduction}

From hate speech and misinformation to verbal violence and death threats, social platforms like X (formally Twitter) have become a favorable space for content that can cause harm through online aggression~\cite{Mansur2023}.
Harmful content is no longer perceived only as a form of immorally expressed opinions but as a recognized global danger that must be prevented for the mental, emotional, and physical safety of online content consumers~\cite{Williams2019}.
It is necessary to be able to detect the sources that generate such toxic behavior on the Internet and to reduce the influence they have. 
The more the spread of hate-speech posts is decreased, the more users will be saved from emotional and psychological damage.
To address these issues, we propose \textsc{StopHC}, a harmful content detection and mitigation architecture for social media platforms.

The main objectives of this paper are: 
(1) to detect problematic behavior, and 
(2) to minimize the spread of such behaviors.
To detect harmful content, we train multiple deep neural network models using different embeddings that consider the syntax (i.e., word embeddings such as Word2Vec~\cite{Mikolov2013a,Mikolov2013b} and GloVe~\cite{Pennington2014}), context (i.e., transformed embeddings such as BERT~\cite{Devlin2019} and RoBERTa~\cite{Liu2019}), and network information (i.e., Node2Vec~\cite{Grover2016}).
Using these approaches, we aim to improve the \textsc{StopHC} detection module by employing the model that better understands both textual content and network structures.
To minimize the spread of such behaviors, we employ 3 different immunization strategies: 
(1) na\"ive (i.e., Highest Degree~\cite{Logins2019}), 
(2) pro-active (i.e., NetShield~\cite{Chen2015NetShield}), and 
(3) contra-active (i.e., DAVA~\cite{Zhang2015DAVA}).
Using these strategies, we aim to improve \textsc{StopHC} mitigation module and offer a graph-dependent solution.

The main contributions of this work are four-fold:
\begin{itemize}
    \item[$C_1$] We propose \textsc{StopHC}, a novel architecture for harmful content detection and mitigation;
    \item[$C_2$] We develop new deep neural network models for harmful content detection;
    \item[$C_3$] We employ immunization strategies to stop the spread of harmful content on social media platforms;
    \item[$C_4$] We perform extensive evaluation testing on two real-world datasets.
\end{itemize}

This work is structured as follows.
Section~\ref{sec:sota} provides insights into the recent literature.
Section~\ref{sec:methodology} presents \textsc{StopHC}'s architecture
Section~\ref{sec:results} offers the experimental evaluation of our solution.
Finally, Section~\ref{sec:conclusions} presents the conclusions of this work and discusses future work.

\section{Related Work}\label{sec:sota}

When dealing with detecting harmful content, the current literature focuses on word embeddings~\cite{Ilie2021}, transformer embeddings~\cite{Petrescu2023,Truica2022Misrobaerta}, sentence transformers~\cite{Truica2022checktaht2022} or document embeddings~\cite{Truica2023}.
Various deep-learning architectures are employed as classifiers for harmful content detection. Notably, most state-of-the-art architectures utilize LSTMs (Long Short-Term Memory networks), GRUs (Gated Recurrent Units), and/or CNNs (Convolutional Neural Networks) as their core building blocks, e.g., \cite{Badjatiya2017deep, Ilie2021, Truica2022Misrobaerta}.
When dealing with network immunization, the current literature proposed multiple solutions that 
analyze the spread of content online~\cite{Petrescu2019, petrescu2023edsaensemble,Truica2021c} and propose either proactive~\cite{Petrescu2021,Apostol2024contain} or contra-active~\cite{Truica2023MCWDST} immunization strategies.
\textsc{StopHC} is a full solution that builds on top of the current literature and proposes a novel deep-learning architecture for harmful content detection and mitigation.

\section{Methodology}\label{sec:methodology}

\begin{figure}[!htbp]
    \centering
    \includegraphics[width=1\columnwidth]{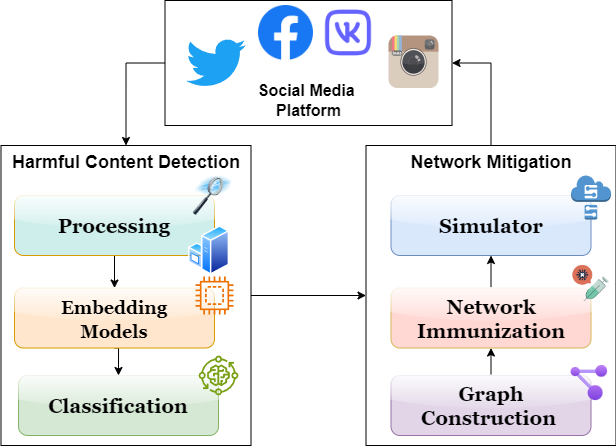}
    \caption{\textsc{StopHC} Architecture}
    \label{fig:architecture}
\end{figure}

In this section, we present \textsc{StopHC}, our architecture for harmful content detection and mitigation.
Figure~\ref{fig:architecture} presents the general architecture for our proposed solution.

To successfully combat the problem of harmful content on social media, we need 2 key components: (1) a detection module, and 
(2) a network immunization module.

The input consists of textual data (e.g., tweets, messages, comments, etc.) and network data collected from social platforms.
The detection module uses textual data to predict if the content is harmful and also marks the users who posted it. 
The network immunization module uses the network data to create a graph based on the user interactions, i.e., the users are nodes, and the interactions are edges (e.g., likes, comments, shares, etc.).
Using this graph, a network immunization algorithm is employed to stop the spread of harmful content. 

\subsection{Harmful Content Detection Module}

This module contains 3 submodules:

\subsubsection{Preprocessing}
The textual data is passed through a data preprocessing pipeline to preserve semantic relations while removing any elements that do not contribute to extracting the context (e.g., hyperlinks, utf8 characters, etc.).
Such elements can confuse the model and lead to irrelevant information being encoded in the embeddings.
Moreover, text cleaning reduces the vocabulary size~\cite{Truica2016a} the model needs to handle and, therefore, increases the model's performance when working with social media data~\cite{Truica2016Cats}.
This standardization allows the model to focus on the core content and generalize better to unseen data. 

\subsubsection{Embedding Models}
Embedding techniques are used to convert textual and network data into vector representations.
For textual data, the embeddings capture the relationships between tokens and encode local and global contexts~\cite{Truica2021}. 
For network data, the embeddings capture the relationships between nodes and the relations between them.
In our implementation, we use 2 word embeddings (i.e., Word2Vec~\cite{Mikolov2013a,Mikolov2013b} and GloVe~\cite{Pennington2014}) and 2 transformer embeddings (i.e., BERT~\cite{Devlin2019} and RoBERTa~\cite{Liu2019}) for textual data, while for network embeddings we use Node2Vec~\cite{Grover2016}.
When both network and textual information are available, we concatenate the embeddings to create the input for the classification module. 

\subsubsection{Classification}
For classification, we train 6 deep neural network models in order to perform extensive ablation testing.
\paragraph{BiLSTM-Dense} This architecture contains a BiLSTM layer followed by two Dense layers, one containing 16 hidden units for flattening the output of the BiLSTM and one for making the predictions using a sigmoid activation function. 
The Dense layer works well with smaller embeddings, where every bit of information is important for making accurate predictions. It acts like a cost-effective filter, extracting the most valuable information for the task.
\paragraph{BiLSTM-CNN-Dense} This architecture adds a CNN layer between the BiGRU layer and the first Dense layer.
\paragraph{BiLSTM-GMP} This architecture replaces the first hidden Dense layer of the BiLSTM architecture with a GlobalMaxPooling (GMP) layer. 
The GMP layer offers a moderate complexity and requires a reasonable number of parameters.
This helps avoid two common problems, i.e., overfitting (the solution would perform poorly on unseen data) and vanishing gradients (the solution would struggle to learn from deeper layers).
This balanced approach makes GMP well-suited for handling larger embeddings, 
allowing it to exploit them better.
\paragraph{BiGRU-GMP} This architecture is similar to the BiLSTM-FMP one, the only difference is the use of GRU units instead of LSTM ones.
\paragraph{BiLSTM-CNN-GMP} This architecture is similar to the BiLSTM-GMP one. The only difference is that a CNN layer is added between the BiLSTM and GlobalMaxPooling (GMP) layers.
\paragraph{2BiLSTM} This architecture adds a second BiLSTM layer to the BiLSTM architecture.

\subsection{Network Immunization Module}

This module also contains 3 submodules:

\subsubsection{Graph Construction}
This submodule constructs the network graph using the users as nodes and the interactions between them as edges. 

\subsubsection{Network Immunization}
It is fair to accept that in this context, we have a limited budget, i.e., the number of nodes to immunize. 
Therefore we must carefully select key nodes for immunization in order to maximize our efforts. 
This module uses the Highest Degree, the NetShield~\cite{Chen2015NetShield}, and the DAVA~\cite{Zhang2015DAVA} algorithms to perform network immunization.
DAVA uses a domination tree to immunize the network, while NetShield detects the spread of toxic information by sorting the nodes using the eigenvalue obtained from the graph's adjacency matrix.
As a baseline, we use the Highest Degree algorithm that selects the top-k nodes with the highest degree.

\subsubsection{Harmful Content Mitigation}
The submodule focuses on the subgraph obtained using toxic nodes, i.e., nodes that spread harmful content, and who to stop the spread of harmful content.
The toxic nodes, detected by the Harmful Content Detection Model, act as seeds for the network immunization algorithm.
Using the results of the immunization algorithm, we plot the spread of harmful content and determine the affected nodes.

\subsection{Graphical User Interface}

\begin{figure}[!htbp]
    \centering
    \includegraphics[width=1\columnwidth]{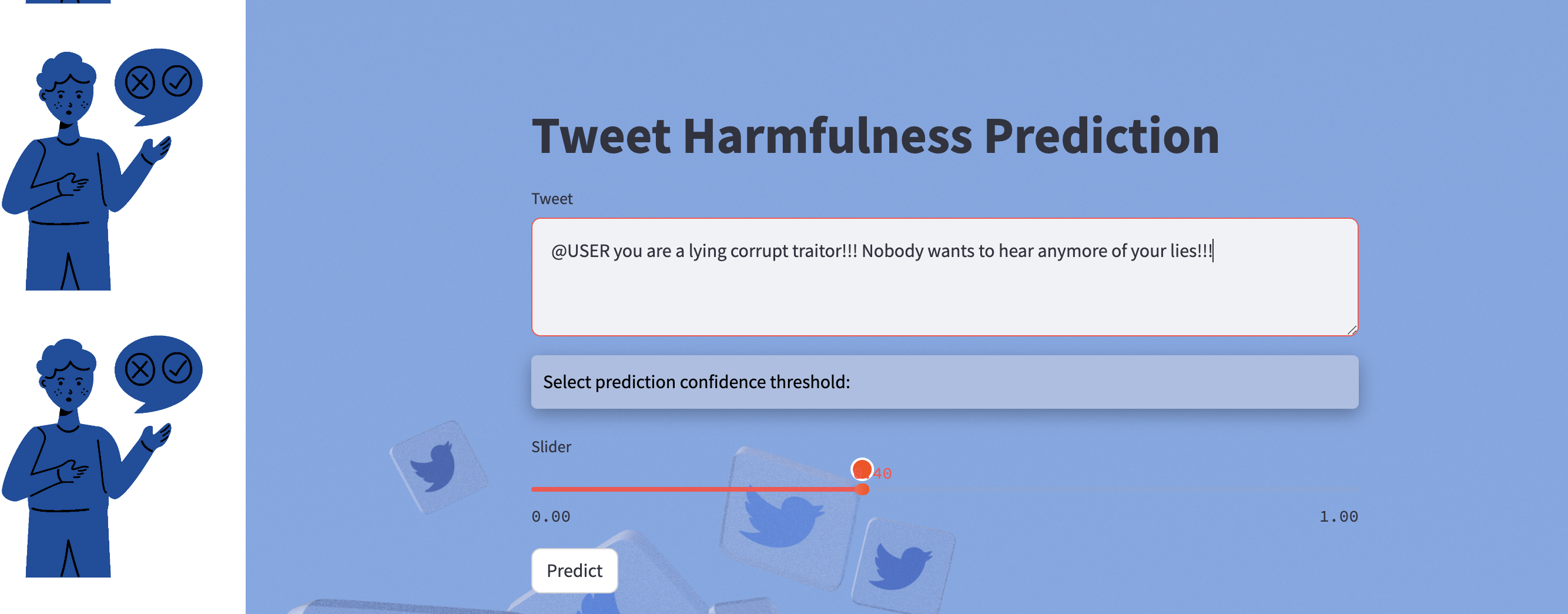}
    \caption{User interface}
    \label{fig:frontend}
\end{figure}

The front-end component (Figure~\ref{fig:frontend}) provides an interface to the user through which she/he can test the \textsc{StopHC} detection solution.
The text entered by the user is directly linked to the preprocessing component of our model. 
The model that offers the best performance is called with the previously found embeddings as a parameter. 
We have also added the possibility of setting a confidence threshold which is going to reflect the boundary of being classified into either harmful or not-harmful categories. 
Finally, the prediction result is displayed in a visually engaging and interactive graphical format.

The code for \textsc{StopHC}’s models and immunization strategies is publicly available on GitHub at https://github.com/DS4AI-UPB/StopHC\_HarmfulContentMitigation.

\section{Experimental Results}\label{sec:results}

To determine the capabilities of our solution \textsc{StopHC}, we propose 2 sets of experiments. 
First, we evaluate the detection module using 2 datasets.
Second, we test the immunization strategies on a real-world dataset.

\subsection{Datasets}

For the experiments, we employ two datasets, i.e., the Hate Speech Dataset~\cite{Davidson2017} and the EXIST2023 Dataset~\cite{Plaza2023exist}.
Hate Speech Dataset~\cite{Davidson2017} contains 33\,458 hate speech, and offensive tweets. 
The number of classes is divided into two, i.e., harmful and not harmful. 
EXIST2023 Dataset~\cite{Plaza2023exist} contains 3\,258 English tweets also classified as harmful and not harmful.
Furthermore, the EXIST2023 Dataset provides network information that we use in our immunization experiments.
As both datasets are imbalanced with a ratio of 1:2 between harmful vs not harmful classes, we use appropriate evaluation metrics that consider this imbalance~\cite{Truica2017}.

\subsection{Classification Results}

For the Word2Vec embedding, we trained our own Skip-Gram model using gensim~\cite{Rehurek2010} with the following parameters: vector\_size = 50 and window = 6.
We used the pre-trained 50-dimension vectors for GloVe.
For the transformer models, we used the pre-trained BERT model \textit{bert-base-uncased} and the pre-trained RoBERTa model \textit{roberta-base-uncased}.
The node2vec embedding is trained on the graph obtained from the EXIST2023 Dataset~\cite{Plaza2023exist}.

To determine the best models, we use 10-fold stratified cross-validation and hyperparameter tuning to obtain the best models for the Hate Speech Dataset~\cite{Davidson2017}. 
When performing hyperparameter tuning, we use grid search with the following parameters:
(1) number of hidden layers $\in \{16, 32, 64 \}$,
(2) optimizers $\in \{Adam, RMSprop \}$, and
(3) learning rate $\in \{0.001, 0.0001, 0.00001 \}$.
The following tables show only the best results after performing hyperparameter tuning and 10-fold cross-validation.

On the Hate Speech Dataset~\cite{Davidson2017}, the best performance is obtained by the model BiLSTM-GMP model with GloVe (Table \ref{tab:model_metrics}), with the GloVe embedding technique, which manages to reach an F1 score of 0.9323., a performance that exceeds the scores recorded by other papers in the State of the Art. 
Glove is able to learn semantic relationships to better fit the proposed task, without overfitting during the training.
Word2Vec achieves comparable performances with the BiLSTM-GMP model as well, by obtaining an F1 score of 0.92.
In comparison with the state-of-the-art models, we observe that our proposed BiLSTM-Dense architecture outperformed them in terms of precision, recall, and F1-score.

\begin{table}[!htbp]
\centering
\caption{Performance Analysis - Hate Speech Dataset \\ (\textit{Note: \textbf{bold} text marks the overall best result})}
\label{tab:model_metrics}
\resizebox{\columnwidth}{!}{
\begin{tabular}{|l|l|c|l|l|l|}
\hline
\textbf{Model} & \textbf{Embedding} & \textbf{Accuracy} & \textbf{Precision} & \textbf{Recall} & \textbf{F1}\\
\hline
BiLSTM-Dense & Word2Vec & 0.9053 & 0.9228 & \textbf{0.9333} & 0.9280\\
BiLSTM-Dense & GloVe & \textbf{0.9122} & 0.9406 & 0.9242 & \textbf{0.9323}\\
BiLSTM-Dense & BERT & 0.8270 & 0.8792 & 0.8883 & 0.8837\\
BiLSTM-Dense & RoBERTa & 0.8255 & 0.9206 & 0.8026 & 0.8576\\
BiLSTM-CNN-Dense & Word2Vec & 0.9107 & 0.9380 & 0.9220 & 0.9315\\
BiLSTM-CNN-Dense & GloVe & 0.9050 & 0.9370 & 0.9143 & 0.9250\\
BiLSTM-CNN-Dense & BERT & 0.8254 & 0.8780 & 0.8850 & 0.8885\\
BiLSTM-CNN-Dense & RoBERTa & 0.8230 & 0.9180 & 0.8000 & 0.8550\\

BiGRU-GMP & Word2Vec & 0.8955 & 0.9252 & 0.8952 & 0.9138\\
BiGRU-GMP & GloVe & 0.8970 & 0.9012 & 0.9056 & 0.9145\\
BiGRU-GMP & BERT & 0.8325 & 0.8359 & 0.9257 & 0.8785\\
BiGRU-GMP & RoBERTa & 0.8671 & 0.8844 & 0.9166 & 0.9002\\
BiLSTM-GMP & Word2Vec & 0.8981 & 0.9551 & 0.8860 & 0.9192\\
BiLSTM-GMP & GloVe & 0.8671 & 0.8844 & 0.9166 & 0.9002\\
BiLSTM-GMP & BERT & 0.8470 & 0.8792 & 0.8883 & 0.8837\\
BiLSTM-GMP & RoBERTa & 0.8683 & 0.8789 & 0.9264 & 0.9020\\
BiGRU-CNN-GMP & Word2Vec & 0.8922 & 0.8982 & 0.9021 & 0.9103\\
BiGRU-CNN-GMP & GloVe & 0.8850 & 0.8950 & 0.9074 & 0.9100\\
BiGRU-CNN-GMP & BERT & 0.8300 & 0.8370 & 0.9205 & 0.8750\\
BiGRU-CNN-GMP & RoBERTa & 0.8650 & 0.8814 & 0.9150 & 0.8975\\

2BiLSTM & Word2Vec & 0.8953 & \textbf{0.9690} & 0.8584 & 0.9148\\
2BiLSTM & GloVe & 0.9053 & 0.9228 & \textbf{0.9333} & 0.9280\\
2BiLSTM & BERT & 0.8505 & 0.8822 & 0.8904 & 0.8863\\
2BiLSTM & RoBERTa & 0.8671 & 0.8873 & 0.9128 & 0.8999\\
2BiLSTM & RoBERTa & 0.8671 & 0.8873 & 0.9128 & 0.8999\\
\hline

Logistic Regression~\cite{Davidson2017} & TFIDF &  N/A & 0.91 & 0.90 & 0.90 \\

Ensemble~\cite{vanAken2018} & Multiple &  N/A & 0.76 & 0.83 & 0.793 \\
BiGRU-Attention~\cite{vanAken2018} & Glove & N/A & 0.77 & 0.82 & 0.790  \\
\hline
\end{tabular}
}
\end{table}

Based on the results obtained on the Hate Speech Dataset~\cite{Davidson2017}, we choose the best-performing architectures, i.e., BiLSTM-Dense, to train on the EXIST2023 Dataset~\cite{Plaza2023exist} (Table \ref{tab:model_metrics2}).
As this dataset also contains network information, we also add the network embedding obtained with Node2Vec.
We observe that when concatenating the work embedding the network embedding, we obtain better results with the transformer models, while the results worsen with the word embedding models. 
This decrease in performance is due to the inability of the models to perform embedding fine-tuning during training. 
Finally, we can confirm that by adding the network's structural information into the detection model, we obtain an improved accuracy for some of our models.
We note that in comparison with the official results from the EXIST2023 challenge, our model obtains a lower F1-score.
This difference is due to Twitter-RoBERTa transformer model that is fine-tuned on the dataset for the specific hate speech detection task.
Also, the F1-score metric does not show how well the model manages to predict hate speech, as it takes into account both precision (which focuses on true positives) and recall (which focuses on true negatives).

\begin{table}[!htbp]
\centering
\caption{BiLSTM-Dense Performance - EXIST2023 Dataset \\ (\textit{Note: \textbf{bold} text marks the overall best result})}
\label{tab:model_metrics2}
\resizebox{\columnwidth}{!}{
\begin{tabular}{|l|l|c|c|c|c|}
\hline
\textbf{Embedding} & \textbf{Node2Vec} & \textbf{Accuracy} & \textbf{Precision} & \textbf{Recall} & \textbf{F1}\\
\hline
Word2Vec & N/A & 0.6994 & 0.5661 & 0.6009 & 0.5830 \\
Word2Vec & Yes & 0.5100 & 0.5161 & 0.2800 & 0.3627 \\
GloVe    & N/A & 0.7147 & 0.5890 & \textbf{0.6096} & \textbf{0.5991} \\
GloVe    & Yes & 0.5577 & 0.4878 & 0.3103 & 0.3797 \\
BERT     & N/A & 0.7117 & 0.5943 & 0.5526 & 0.5727 \\
BERT     & Yes & \textbf{0.7331} & 0.6753 & 0.4561 & 0.5845 \\
RoBERTa  & N/A & 0.7224 & \textbf{0.6767} & 0.3947 & 0.4986 \\
RoBERTa  & Yes & 0.7239 & 0.6600 & 0.4342 & 0.5238 \\
\hline
Twitter-RoBERTa~\cite{Petrescu2024} & No & N/A & N/A & N/A &  \textbf{0.7475} \\

\hline
\end{tabular}
}
\end{table}

\subsection{Immunization Results}

\begin{figure}[!htbp]
    \centering
        \subfigure[Edges of the most influential 10 nodes using NetShield~\label{fig:netshild}]{
        \includegraphics[width=0.48\columnwidth]{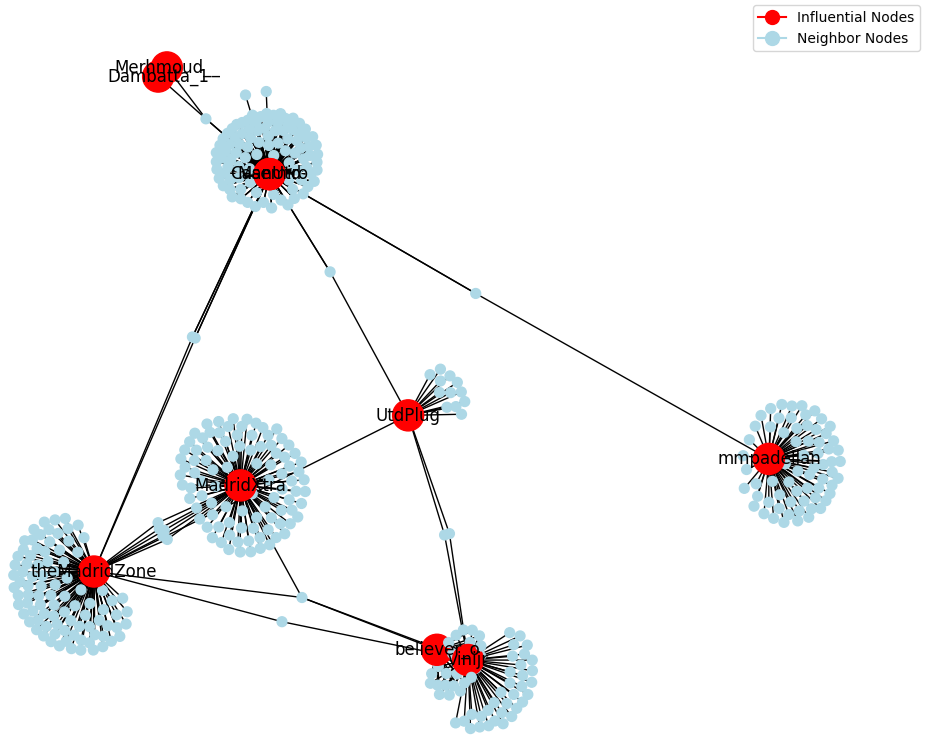}}
        \subfigure[Edges of the most influential 10 nodes using DAVA~\label{fig:dava}]{
        \includegraphics[width=0.48\columnwidth]{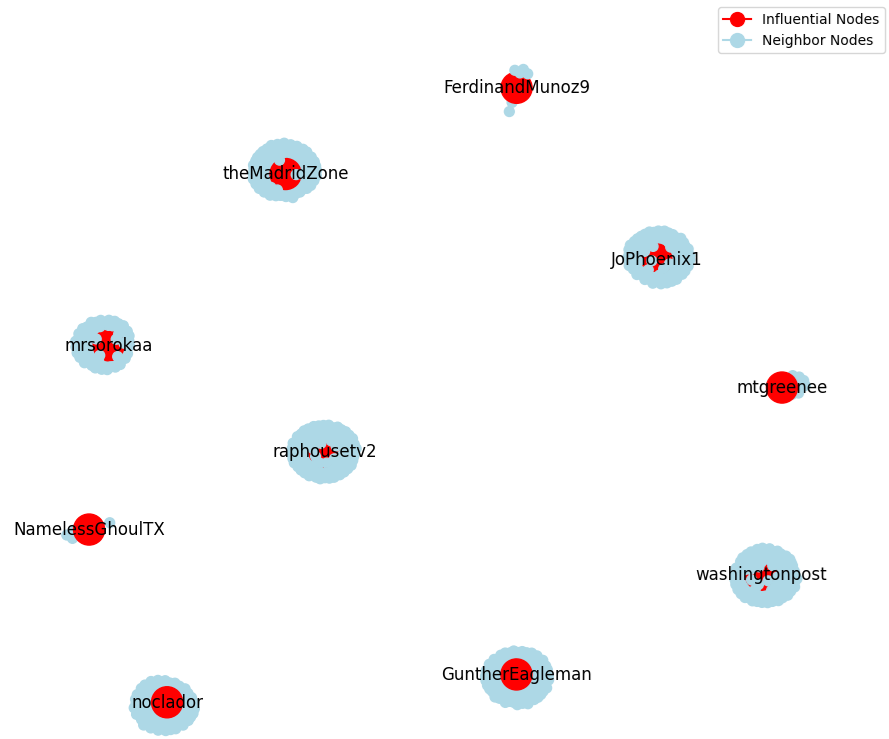}}
    \caption{Edges of the most influential 10 nodes using different immunization algorithms}
    \label{fig:graphs}
\end{figure}

For the network immunization results of \textsc{StopHC}, we use DAVA and NetShield.
In our comparison, we use the Highest Degree algorithm as a baseline. 
Figure~\ref{fig:graphs} shows the links between the nodes selected as toxic by the harmful Content Detection module and their direct neighbors obtained by the Graph Construction submodule.
The goal of these experiments is to determine the position of the toxic nodes in the graph and whether or not they are an element of connectivity between several clusters that spread harmful content.

We observe that NetShield (Figure~\ref{fig:netshild}) provides as an output a set of nodes that balance both centrality and connectivity.
We expect the nodes selected by Netshield to be at the bridge between two clusters.
By deleting these nodes, the overall connection of the graph is diminished.

On the other hand, DAVA does not manage to offer a global immunization strategy, but rather a local one that might be produced by certain nodes. 
Thus, DAVA does not manage to detect toxic nodes that represent bridges between several clusters.

\begin{figure}[!htbp]
    \centering
    \includegraphics[width=1\columnwidth]{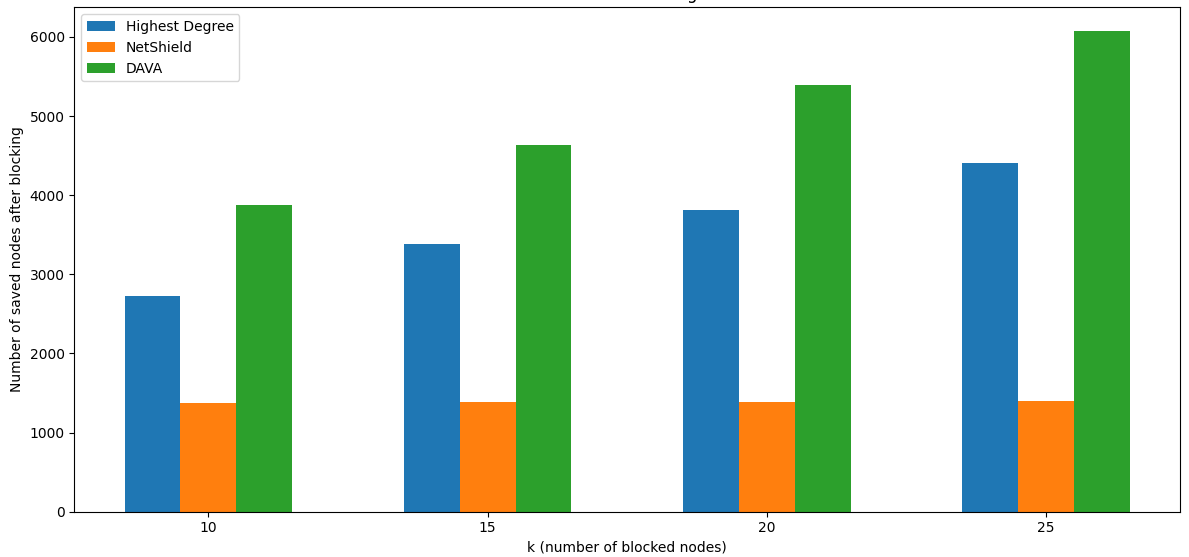}
    \caption{Number of saved nodes}
    \label{fig:nodes_saved}
\end{figure}

Figure~\ref{fig:nodes_saved} presents the number of saved nodes using the different immunization strategies.
DAVA manages to immunize more nodes because it specializes in handling local spreads.
At the same time, we observe that immunizing the most popular nodes would also prevent the propagation of the content, as they lie along the path of the infected nodes in our subgraph. 
Netshield fails to save as many nodes as the other two algorithms due to the structure of the graph, which does not favor the perception of vulnerability through the detection of eigenvalue.

\begin{table}[!htbp]
\centering
\caption{Execution time for each algorithm (in seconds)}\label{tab:time}
\begin{tabular}{|c|c|c|c|c|}
\hline
\textbf{Algorithm} & \textbf{k=10} & \textbf{k=15} & \textbf{k=20} & \textbf{k=25} \\ \hline
Highest Degree & 0.3253 & 0.3131 & 0.3161 & 0.3010 \\ 
NetShield & 1.5244 & 1.9489 & 1.7003 & 1.7489 \\ 
DAVA & 7.3669 & 10.4963 & 13.0861 & 15.4283 \\ 
\hline
\end{tabular}
\end{table}

To determine the scalability of our solution, we also perform scalability testing and record the execution times (Table~\ref{tab:time}), where $k$ is the number of nodes to immunize, i.e., the budget.
We observe that DAVA has a higher execution time than Highest Degree and NetShield.

\section{Conclusions}~\label{sec:conclusions}

In this paper, we propose \textsc{StopHC}, a harmful content detection and mitigation architecture for social media platforms.
Using \textsc{StopHC}, first, we detect harmful content employing deep neural network detection models, and second, we perform network immunization to stop the spread of harmful content online.
The experimental validation shows that our solution manages to accurately detect harmful content, stop the spread of toxic information online, and scale linearly with the number of nodes we want to immunize. 
Furthermore, the use of network embeddings improves the accuracy of the content detection models that employ transformed models.

In future work, we aim to improve the network immunization module by using SparseShield~\cite{Petrescu2021}, CONTAIN~\cite{Apostol2024contain}, and MCWDST~\cite{Truica2023MCWDST}.
We also aim to use novel architectures for detecting harmful content such as DANES~\cite{Truica2023danes} or MisRo{\AE}RTa~\cite{Truica2022Misrobaerta}.

\section*{Acknowledgment}

This work is supported in part by
(1) 
The National University of Science and Technology Politehnica Bucharest through the ``PubArt'' project.
(2) The German Academic Exchange Service (DAAD) through the project ``iTracing: Automatic Misinformation Fact-Checking'' (DAAD grant no. 91809005); and
(3) The Academy of Romanian Scientists through the funding of ``SCAN-NEWS: Smart system for deteCting And mitigatiNg misinformation and fake news in social media''.

\bibliographystyle{IEEEtranS}
\bibliography{bibliography}

\end{document}